# A Scoping Review of Energy Load Disaggregation

Tolnai, Balázs András; Jørgensen, Bo Nørregaard; Ma, Zheng Grace

Go to publication entry in University of Southern Denmark's Research Portal

Corresponding author: Balázs András Tolnai

# A Scoping Review of Energy Load Disaggregation

Balázs András Tolnai[1[0009-0004-4183-4340]] and Zheng Ma[2[0000-0002-9134-1032]] and Bo Nørregaard Jørgensen[3[0000-0001-5678-6602]]

[123] SDU Center for Energy Informatics, the Maersk Mc-Kinney Moller Institute, University of Southern Denmark, Odense 5230, Denmark
`bat@mmmi.sdu.dk`

**Abstract.** Energy load disaggregation can contribute to balancing power grids by enhancing the effectiveness of demand-side management and promoting electricity-saving behavior through increased consumer awareness. However, the field currently lacks a comprehensive overview. To address this gap, this paper conducts a scoping review of load disaggregation domains, data types, and methods, by assessing 72 full-text journal articles. The findings reveal that domestic electricity consumption is the most researched area, while others, such as industrial load disaggregation, are rarely discussed. The majority of research uses relatively low-frequency data, sampled between 1 and 60 seconds. A wide variety of methods are used, and artificial neural networks are the most common, followed by optimization strategies, Hidden Markov Models, and Graph Signal Processing approaches.

## 1 Introduction

The increasing electricity demand coupled with the uneven production of renewable energy sources and imbalanced electricity consumption puts immense pressure on the existing power grids [1]. Demand-side management and load shifting can address these issues by redistributing electricity consumption away from peak periods, thus creating a more balanced consumption curve or aligning consumption peaks with electricity generation peaks [2, 3]. Accurate knowledge of individual appliance consumption at specific times can help identify shiftable loads, enabling more precise demand-side management. Load disaggregation can also promote electricity-saving behaviors among individuals by raising awareness of their consumption patterns [1-2].

Load monitoring, or load disaggregation, involves measuring the electricity consumption of individual consumers. Intrusive Load Monitoring (ILM) is a traditional method that requires submetering appliances, necessitating a measurement device for each monitored consumer [4]. This approach results in high hardware requirements, which can be expensive and logistically challenging to deploy. In contrast, Non-Intrusive Load Monitoring (NILM) was first introduced by [5]. NILM calculates the consumption of individual appliances by using the electricity measurements already available at aggregation points, such as electricity meters.

However, the field of energy load disaggregation still lacks a comprehensive overview of the various domains, datasets, and methodologies employed in the research. Without a comprehensive overview, it becomes challenging for researchers and practitioners to identify knowledge gaps, potential areas for innovation, and best practices for effective load disaggregation [6]. Therefore, this paper aims to present an in-depth review of load disaggregation literature using a scoping review method, providing a systematic analysis of the current state of the field, and identifying challenges, and opportunities for future research and development.

The paper is organized as follows: First, the methodology section presents the scoping review process including the search strategy and inclusion and exclusion criteria; the results section provides an overview of the key findings from the reviewed literature, highlighting the domains, datasets, and methods used in energy load disaggregation research. Subsequently, the discussion section delves into the implications of these findings, addressing the challenges, limitations, and opportunities within the field, as well as offering recommendations for future research directions. Finally, the conclusion section summarizes the main contributions of the paper and emphasizes its relevance to the ongoing development of load disaggregation research and applications.

## 2  Methodology

This paper employs the scoping review methodology introduced in [7] which involve a systematic literature search to identify and analyse the relevant articles on load disaggregation. The first step in the literature search process was to apply the search string "load disaggregation" in the Web of Science database. The reason for choosing the Web of Science as the primary source for this search was due to its comprehensive coverage of high-quality, peer-reviewed research articles from various disciplines, including energy and engineering fields. No filtering was applied based on publication date. The initial search resulted in 131 articles with the search string appearing in their titles. The reason for focusing on titles was to ensure that the selected articles had a strong and direct focus on load disaggregation as a primary research topic, which is essential for the scope of this review.

To further refine the search and ensure that the articles matched the paper's scope, the search string "energy OR electricity OR heat" was applied to the abstracts of the 131 identified articles. This step was necessary to confirm that the selected articles addressed load disaggregation in the context of energy, electricity, or heat. This filtering process resulted in 122 relevant articles for further analysis.

To ensure the quality of the literature analysis, only peer-reviewed journal articles were considered for the review. A full-text search was conducted, and full texts for 72 out of the 122 relevant articles were retrieved. The 72 full-text articles were thoroughly reviewed, and relevant data, such as authors, publication year, research methods, main findings, and key contributions, were extracted.

## 3 Results

The scoping review results of load disaggregation research can be divided into the applied domains, data and data sources, as well as the various methods employed.

### 3.1 Applied Domains

As illustrated in Fig. 1, residential homes and households represent the most common domain, with 54 out of the 72 investigated articles focusing on disaggregating the load for these types of consumers. Other domains explored include commercial buildings (4 articles), electric vehicles (3 articles), and industrial parks (2 articles). Three articles examined the disaggregation of larger, bulk supply points for multiple buildings.

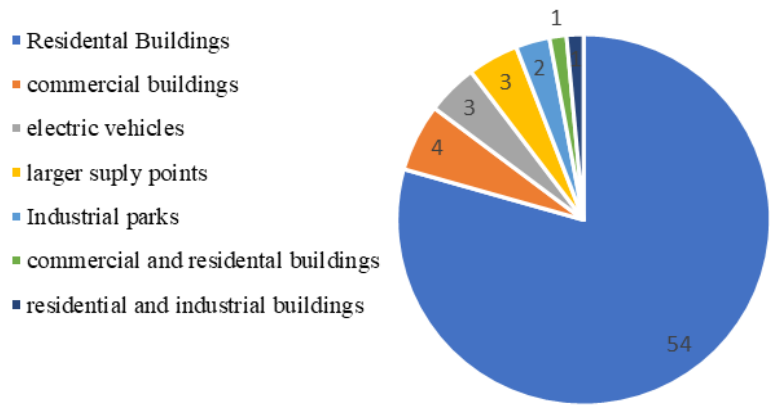

**Fig. 1.** The investigated domains in the literature

Households are the predominant research domain in load disaggregation. Most studies in this area aim to disaggregate major, easily metered household appliances, such as washing machines, dishwashers, kettles, microwaves, and ovens [8-15]. Four articles [2, 16-18] specifically focused on identifying heating and cooling-related loads and separating them from the rest. Two articles aimed to disaggregate household electricity load while incorporating electricity generated by solar panels [19, 20]. Another study sought to predict battery failures by adding battery simulations to the load profile and detecting anomalous battery behaviour through load disaggregation [21].

In the commercial sector, two studies examined the measurement of typical office devices like lamps, computers, and coffee machines [22]. One of them [23] used disaggregation to identify individuals' departure and arrival times. An article [24] disaggregated the lighting load profile using only ambient lighting information measured by light sensors and knowledge of the light bulb numbers and energy uptake, while [25] employed NILM to disaggregate various cooling loads from an aggregate cooling load already separated from the overall electricity consumption.

Research in the industrial sector primarily focused on disaggregating the consumption of different industrial tools. Disaggregation of various tools in [26] was used to enhance a demand response scheduling algorithm's performance. Similarly, [27] employed load disaggregation to assist demand-side management.

Three studies concentrated on electric vehicle charging. In [28] a framework was developed for disaggregating the total energy of plug-in electric vehicles at the feeder head level. A different article [29] modelled the electricity consumption of charging stations, using similarities between the created models to better understand the future requirements of similar charging stations. Article [30] disaggregated feeder head-level consumption into separate smart meter readings, encompassing both industrial and residential buildings. An article [4] used bulk supply point measurements to disaggregate them into several categories, such as switch mode power supply, different induction motors, lighting, rectifiers, and resistive load, while [8] also considered plug-in electric vehicles at the feeder head level.

### 3.2 Data and Data Sources

Load disaggregation research primarily focuses on electricity loads, occasionally incorporating other information such as outdoor temperature and solar radiation. One article [27] also includes gas consumption as supplementary information for electricity load disaggregation. Multiple publicly available datasets are commonly used in the literature, as demonstrated by the distribution in Fig 2. The most popular dataset is the REDD (Reference Energy Disaggregation Data Set), used in 16 standalone articles and an additional 16 articles in conjunction with other datasets. The UK-DALE (UK Domestic Appliance-Level Electricity) dataset is the second most popular, used three times alone and 11 times with other datasets. The AMPds (Almanac of Minutely Power dataset) ranks third, appearing in seven articles [31-34]. The least popular datasets—LIFTED, GreenD [35], Eco [35], Low Carbon London (LCL) [36], and Rainforest Automation Energy Dataset (RAE) [37]—each appeared only once in the reviewed articles. Furthermore, multiple studies used non-public datasets.

**Residential buildings.** The most commonly used dataset for residential household research is the REDD dataset, featured in 29 articles, e.g., [21, 38-45], followed by the UK-DALE dataset, used in 12 articles, e.g., [39, 46-48]. The least used datasets, such as LIFTED, appeared in only one article [49]. Meanwhile, most residential load disaggregation research is conducted on relatively low-frequency data, as seen in 37 articles. For example, [50] tested their model on data sampled at 15, 30, and 60-second intervals, while [51] used data sampled at 6 and 60-second intervals. Six articles use data between 1 and 15 minutes [16, 52, 53], five use frequencies higher than 1 second [49, 54], two work with hourly data [2, 55], and one with 30-minute sampling rate [17].

The REDD dataset, publicly available and comprising metering data from real US residential houses, monitors central, aggregate consumption at a high frequency of 15 kHz, while individual appliances are monitored at lower frequencies of 0.5 or 1 Hz [56]. The UK-DALE dataset, collected from UK domestic houses, records central consumption at a high frequency of 16 kHz (downsampled from 433 MHz) and appliance-level consumption at a lower frequency of 1/6 Hz, including up to 54 appliances per house [15].

Regarding frequency, most research (42 articles, such as [50] and [21]) is based on a range of 1 minute to 1 second. Six articles work with 1 to 15 minutes, three with a 30-minute sampling rate, and seven with an hourly sampling rate. An additional seven articles utilize high-frequency data [57-59], ranging from 60 measurements per second up to 44.1 kHz. This information is depicted in Fig. 3.

**Commercial buildings.** Research on commercial office buildings typically relies on non-public datasets. One study [23] utilized data from a preliminary study conducted in an office space owned by the University of Nebraska. Another study focusing on heating and cooling loads in an office environment collected data from a large Chinese office (32,000 m2) with hot summers and cold winters [25].

In terms of sampling rates for commercial buildings, three articles use 1 second to 1-minute rates, though only one employs NILM [23]. The other two utilize hourly sampling rates [25, 60]. Industrial research is predominantly based on private datasets, with articles [26, 27] using data collected from Iranian industrial zones.

The WHITED dataset [61], employed by [62, 63], includes both industrial and residential appliances with high measurement frequency (44.1 kHz), while [26, 27] use real industrial data applying hourly sampling rates.

**Electric vehicles**. There are limited datasets involving electric vehicles. The EA Technology company conducted the My Electric Avenue project in 2017 to understand the impact of electric vehicles on local electricity grids. The project released a dataset used by article [28]. Another British study, the Low-Carbon-London project, also released electric vehicle charging data, utilized by article [36].

The three articles in this domain used varying sampling rates. For residential plug-in electric vehicle load disaggregation, the article employed hourly data [28], while [36] used a 30-minute sampling rate in their research. In article [28], which focuses on feeder head-level electric vehicle load disaggregation, the electric vehicle data was combined with zonal load data released by Independent System Operators in New England. Duke Energy provided a dataset for [30], collected in North Carolina.

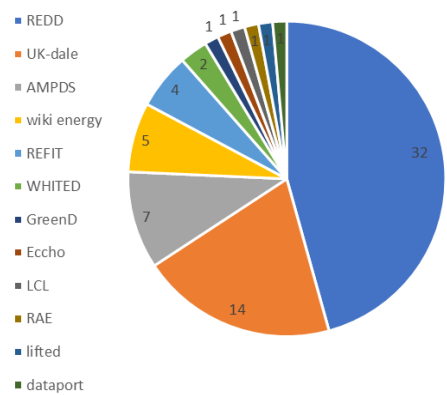

**Fig. 2.** The distribution of the public datasets used.

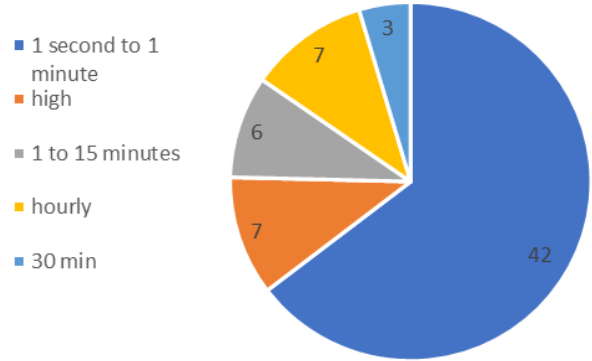

**Fig. 3.** Data frequency distribution

### 3.3 Related Methods

Numerous methods and approaches exist for NILM, which can be primarily classified into supervised and unsupervised learning algorithms [2]. Supervised learning algorithms rely on labeled data for training, which can be difficult to obtain as it necessitates intrusive load monitoring techniques. Consequently, several studies focus on unsupervised models that do not require a learning phase, although manual tuning is necessary [64].

Many NILM methods are event or state-based, making the feature extraction and event detection phases crucial for these algorithms. As illustrated in Fig. 4, neural network-based solutions are the most prevalent due to the wide variety of models available. Long Short-Term Memory (LSTM) models are popular as they can capture the temporal features of the data. For example, [46] employs a multiple-output LSTM model, while [65] applies a deep composite LSTM network to address the disaggregation problem. Convolutional Neural Networks (CNN) are the second most common neural network in the literature. Article [39] investigates the effect of attention mechanisms on both CNN and LSTM networks, while [25] explores the influence of non-electric data on load disaggregation by testing three neural network models, including a sequence-to-point and sequence-to-sequence version of CNN, and a denoising autoencoder.

The second most common approach defines disaggregation as an optimization problem. Optimization-based methods generally do not rely on events [66]. Several optimization algorithms exist, such as integer programming used by [67, 68], an improved version of the Prey-Predator Optimization Algorithm in [50], and particle swarm optimization in [62, 69]. Graph Signal Processing (GSP) is another popular method and can be employed in both supervised and unsupervised settings, e.g., [70-72]. For instance [16] apply a supervised GSP algorithm on real data sampled at 15-minute intervals, while Ming-Yue Zhai also uses GSP in an article [36].

Hidden Markov Models (HMM), often factorial HMMs, are frequently used either independently or in conjunction with other methods. Typically employed in a non-event-based manner, HMMs can also be utilized based on events to address issues related to other HMM algorithms, as demonstrated in [49]. Wen Fan [54] uses FHMM to model appliance states and an optimization method to solve the disaggregation problem. Less common methods include probability-based approaches, such as Gaussian Mixture Models in [64], or Linear Discriminant Classifier Group in [73].

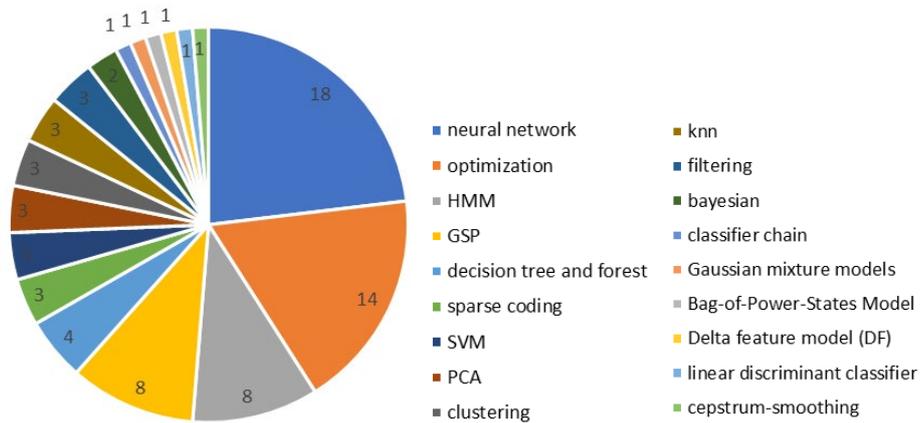

**Fig. 4.** The distribution of methods used for Load disaggregation.

**Applied Domains**. In the residential domain, the distribution of models aligns with the overall results. The four most common model types are neural networks, optimization and HMM-based models, graph signal processing, while less popular methods include Gaussian mixture models and classifier chains [74].

For office buildings, three studies utilize NILM. One study employs random forest and Fourier transform for feature extraction [25]. Article [60] works with both residential and commercial buildings using a mask-based deep neural network to disaggregate flexible, primarily heating and cooling-related, electricity loads [75]. In the third article, Hamed Nabizadeh Rafsanjani develops a disaggregation framework that uses the DBSCAN clustering method to identify electricity load events caused by the same person.

Both research articles conducted on industrial parks employ optimization frameworks. Article [26] utilizes an algorithm called OLDA (Optimization-Based Load Disaggregation Algorithm) developed in [39]. Another article creates an optimization model that leverages industry-specific features alongside electricity load for improved disaggregation results [27].

Articles working with electric vehicles employ various methods. A Kalman filter-based method is used in [28], which disaggregates feeder head-level residential electric vehicle load. Another study that attempts to identify the number of electric vehicles at a charging station employs matching pursuits, a sparse approximation algorithm [36].

**Data and Data Sources.** The most researched sampling frequency ranges between 1 second and 1 minute. These studies employ a wide variety of methods, with neural networks being the most common. For example, some articles explore the use of attention mechanisms to improve disaggregation models [39, 51, 76]. The second most popular approaches are HMM-based [77, 78] and GSP-based models [36, 43]. Less commonly used methods include filtering techniques, such as particle filtering in article [79], and decision trees [47].

Optimization frameworks are the most utilized methods for data with frequencies between 1 and 15 minutes, such as linear optimization used by [67]. Other less common methods include neural networks [52] and clustering [18]. For very low-frequency data, sampled at 30 minutes or an hour, most studies employ neural networks [17] or optimization [26]. Less frequently used are GPS [55] and sparse coding [36]. For high-frequency data, optimization [62] and HMM-based methods [49] are the most common. Overall, the diverse range of methods employed across different domains and data frequencies demonstrates the adaptability and versatility of NILM techniques in addressing various load disaggregation problems.

## 4 Discussion

Among the five main applied domains, residential households represent the most researched area and have the most freely available datasets, leading to the development of new models specifically for residential data. Related challenges include the difficulty of gathering high-frequency data and the limited computational capacity of smart meters, which restricts the integration of disaggregation algorithms [80]. Research on bulk and feeder head level data is less common, partly due to the limited availability of data and the impossibility of detecting individual appliance state changes at the feeder head.

The challenge related to commercial buildings is mainly the low availability of data [81, 82]. Office buildings typically have numerous appliances, which would be expensive to individually submeter [83]. A significant portion of consumption comes from lighting, which is more difficult to submeter than plug-in devices [84].

The main challenge in research on industrial data faces challenges such as industrial secrecy, which limits the availability of public datasets, and the difficulty of generalizing between different industrial consumers due to variations in industrial processes [85]. Submetering all devices to collect disaggregation data is also challenging.

Electric vehicle research faces challenges due to the diverse electricity load that electric vehicles produce, as well as the added complexity introduced by Vehicle-to-Grid (V2G) capable vehicles feeding electricity back into the grid [86, 87].

Regarding data and data sources, studies most commonly use data with a frequency range of 1 second to 1 minute, which is still considered low-frequency data and does not permit the extraction of transient features. Consequently, steady-state features are used for load classification, which can be more challenging to distinguish.

High-frequency data allows for the extraction of transient states, which are easier to classify than steady states. However, obtaining and processing high-frequency data is more difficult and expensive due to the need for specialized equipment and increased computational resources.

Working with lower sampling rate data complicates the extraction of even steady-state transitions, making disaggregation particularly difficult for appliances with short operation times and low power consumption. To address this issue, complementary information is often employed, such as outside temperature for heating and cooling loads or occupant behavior for residential loads.

Various methods are used for load disaggregation, with the primary categories being unsupervised and supervised learning. Supervised learning algorithms can be accurate but require large amounts of data, which can be difficult to acquire. Unsupervised algorithms are often less accurate and cannot identify specific appliances without prior knowledge. Neural networks, optimization GSP, and HMMs are among the popular methods employed. However, each method has its limitations, such as computational complexity, sensitivity to noise, inability to handle unknown devices, and the need for high-frequency data.

Considering the multiple obstacles in collecting high-frequency data, future research should focus on low-frequency disaggregation. The development of computationally inexpensive yet accurate algorithms is essential, especially for residential domains, to achieve smart meter integrable solutions.

Exploring lightweight state-of-the-art deep learning approaches, such as knowledge distillation, and investigating the potential of federated learning to overcome privacy and secrecy issues could lead to models with better generalization capabilities. Moreover, expanding research into fields other than domestic loads and exploring disaggregation of other energy carriers, such as gas or district heating, can open up new opportunities for energy management and efficiency.

## 5  Conclusion

This paper presents a comprehensive review of energy load disaggregation, offering an overview of the field by examining the domains of research, datasets employed, and methodologies utilized. The most extensively investigated area in the reviewed literature involves the disaggregation of the electricity load in residential homes, with a focus on individual appliances. Industrial load disaggregation and the disaggregation of electricity at bulk supply points remain the least explored topics.

The majority of studies investigate sampling frequencies between 1 second and 1 minute, while 1 to 15-minute and 30-minute sampling rates receive less attention. In terms of methods, deep neural network models are the most widely used due to their versatility, followed by optimization-based approaches, Hidden Markov Model (HMM) based techniques, and graph signal processing. Less commonly explored methods include Gaussian mixture models, filtering, and delta feature models.

This paper's scope is limited to the publications identified by the utilized search string, which may exclude relevant studies. Future research should employ search strings with broader coverage, such as "non-intrusive load monitoring." As this is a literature review paper, the recommendations provided require further investigation and validation through empirical research.

Future research should prioritize the reduction of computational requirements for load disaggregation, as current methods often entail high computational costs. Efforts should also be made to improve the accuracy of frameworks working with very low-frequency data, thereby easing data collection requirements and increasing the accessibility of such models. Additionally, currently under-researched areas, such as industrial load disaggregation, warrant further exploration.

## Acknowledgements

This paper is part of the project "Data-dreven smarte bygninger: data sandkasse og konkurrence" (Journalnummer: 64021-6025) by EUDP (Energy Technology Development and Demonstration Program), Denmark.